\documentclass[]{emulateapj}
\usepackage{graphicx, amsmath, amsthm, amssymb,color}

\newcommand{\ba}{\begin{eqnarray}}
\newcommand{\ea}{\end{eqnarray}}

\newcommand{\out}{_\mathrm{out}}
\renewcommand{\in}{_\mathrm{in}}
\newcommand{\oct}{_\mathrm{Oct}}
\newcommand{\nvec}{\hat{\mathbf{j}}}
\newcommand{\uvec}{\hat{\mathbf{e}}}

\newcommand{\evec}{\mathbf{e}}
\newcommand{\jvec}{{\mathbf{j}}}

\newcommand{\inout}{_\mathrm{in-out}}

\def\red#1 {\textcolor{red}{#1}\ }   
\def\blue#1 {\textcolor{blue}{#1}\ }   

\shorttitle{White dwarf pollution after planetary engulfment}
\shortauthors{Petrovich \&  {Mu{\~n}oz}}

\begin{document}

\title{Planetary engulfment as a trigger for white dwarf pollution}
\author{Cristobal Petrovich\altaffilmark{1,2} \& Diego J. {Mu{\~n}oz}\altaffilmark{3}}
\altaffiltext{1}{Canadian Institute for Theoretical Astrophysics, University of Toronto, 
60 St George Street, ON M5S 3H8, Canada; cpetrovi@cita.utoronto.ca}
\altaffiltext{2}{Centre for Planetary Sciences, Department of Physical \& 
Environmental Sciences, University of Toronto at Scarborough, Toronto, 
Ontario M1C 1A4, Canada}
\altaffiltext{3}{Cornell Center for Astrophysics and 
Planetary Science, Department of Astronomy, Cornell University, Ithaca, NY 14853, USA}

\begin{abstract}
The presence of a planetary system 
can shield a planetesimal disk 
from the secular gravitational 
perturbations due to  distant outer massive objects (planets or
stellar companions).
As the host star evolves off the main sequence to become a 
white dwarf, these planets can be engulfed during the giant phase, 
triggering secular instabilities and leading to the tidal disruptions
of small rocky bodies. 
These disrupted bodies can feed the white dwarfs
with rocky material and possibly explain the
high-metallicity material in their atmospheres.
We illustrate how this mechanism can operate when the
gravitational perturbations are due to the KL mechanism
from a stellar binary companion, a process that is 
activated only after the planet has been removed/engulfed.
We show that this mechanism can explain the observed 
accretion rates if: 
(1) the planetary engulfment happens
fast compared to the secular timescale, which is generally
the case for wide binaries ($>100$ AU) and planetary engulfment 
during the Asymptotic Giant Branch ;
(2) the planetesimal disk has a total mass 
of $\sim10^{-4}-10^{-2}M_\oplus$.
We show that this new mechanism can provide a steady
supply of material throughout the entire life 
of the white dwarfs for all cooling ages and
can account for a large fraction (up to nearly half) 
of the observed polluted WDs.

\end{abstract}                                     

\section{Introduction}
\label{sec:intro}

Atmospheric metals are not expected to be 
present in isolated white dwarfs (WDs) with effective temperatures 
below $\sim25,000$ K.  At these temperatures, radiative forces become too weak 
 \citep{chayer95} to significantly counteract the quick gravitational settling that
sinks material heavier than helium in extremely short timescales 
compared to the typical cooling ages of WDs \citep{FM79,K09}.
However, it has been found that $\sim25\%-50\%$ of all field
WDs exhibit spectral lines that are indicative of the presence
of metals in their atmospheres \citep{Z03,Z10,K14}.

The high-metallicity material found in the 
atmospheres of most of these ``polluted'' WDs
is consistent with the composition of rock-forming
material \citep{Z07,G12,farihi13,JY14}.
This observation 
suggesting that pollution 
comes from minor rocky bodies (e.g., asteroids).
One possibility is that these rocky bodies get
very  close to the WD so they can be tidally disrupted
and then accreted.
Further support of this picture comes from observations
of circumstellar disks --revealed by infrared excess in the stellar spectrum--
around many polluted WDs (see \citealt{farihi16} for a recent review). These disks
orbit within ${\sim}1R_\odot$, roughly the distance
at which the material would reside after the tidal disruption (the Roche radius).
All the WDs with detected disks have atmospheric pollution.
More recently, this picture has been reinforced by the recent 
observation of minor bodies
transiting the polluted WD 1145+017
\citep{vanderburg15,alonso16,G16,rap16,xu16}.

Although the leading explanation for WDs pollution
--the accretion of tidally disrupted asteroids--
seems robust and well supported by observations,
the underlying dynamical mechanism responsible for
placing these rocky bodies in star grazing orbits
remains much less constrained and understood.
A better understanding of this mechanism 
can lead to new insights into initial conditions leading
to WD pollution, as well as into the
long-term dynamics and evolution
of the planetary systems around  WDs
and/or their progenitors (typically A and F stars; see
\citealt{veras16} for a recent review on this subject). 

A theoretical model to explain the WD pollution 
from planetary dynamical  instabilities
was put forward by \citet{DS02}.
According to their model, a planetary system that is marginally stable
throughout the main sequence can become unstable due to stellar 
mass loss during post-MS evolution. This global
instability can then promote
some asteroids into star grazing orbits.
This idea has been explored in more detail using 
realistic numerical $N$-body integrations of multi-planet 
systems (no asteroids) and stellar evolution 
\citep{veras13,MVV14,VG15}.
Similarly, the mass loss of the host star can widen the region 
around mean-motion resonances where chaotic diffusion
of asteroids acts efficiently, leading to their posterior
tidal disruption \citep{B11,DWS12,FH14}. As well,
mass loss in close binary systems can drive the outermost planetesimals into the chaotic orbits, with one of the possible outcomes being collisions with either one of the stars \citep{kratter12}.

Thus far, these proposed dynamical mechanisms rely 
on generally short-timescale instabilities
(either scattering or mean-motion-resonance overlap) triggered (or enhanced) by mass
loss or simply by the aging of the planetary systems, and
still face some difficulties. In particular, these mechanisms
are subject to the following constraints:
\begin{enumerate}

\item {\it the delivery of material 
must happen for WDs of all ages.}

The observations seem to show that neither the rate of polluted WDs, 
nor that the level of pollution decreases with the WD
cooling age \citep{KGF14,wyatt14}. Thus,  to explain the observed pollution rate,
the underlying mechanism should be able to deliver enough material into the WD's atmosphere
independently of how much time it has passed since the stellar mass loss
phase.

\item {\it The supply of material into white dwarf grazing orbits
must be a steady process.}

Both the large observed rate of polluted WDs and 
the short timescales that follow a disruption event (or order
the orbital timescale) require of a sustained process
to deliver bodies toward disruption. The formation of a debris
disks following disruption can extend the duration of the delivery 
toward the stellar atmosphere, but its associated
timescale is still short compared to the 
cooling ages of most polluted WDs  \citep{veras14b,veras15}.

\item {\it The reservoir of rocky material has to be long-lived.}

The amount of material waiting to be delivered toward the star 
cannot be arbitrarily large.
A planetesimal disk can be destroyed by 
a collisional cascade, shattering
the rocky bodies down to dust, which can be
blown out during the RG and AGB phases
by radiation pressure (e.g., \citealt{BW10}).
All else being equal, disks with lower surface densities and
at larger separations can survive for longer timescales,
possibly avoiding this fate (e.g., \citealt{wyatt07,HT10,BW10}). 

\end{enumerate}

In this paper, we propose a new mechanism that overcomes
(or at least alleviates) these difficulties. 

We propose that the nature of the instabilities, which
drives the material in a planetesimal disk into disrupting
orbits, is secular (not scattering nor driven by mean-motion 
resonances) and that the instabilities are initiated only at the very 
end of the the stellar evolution (AGB phase) once a stabilizing,
pre-existing planetary system is engulfed by an extended stellar envelope.
This mechanism can provide steady pollution over all ages of the 
WD (overcoming the difficulties 1 and 2), while working for a low surface density
disk that remains dynamically cold during the main sequence, and that gets
gradually depleted long after mass loss has taken place (addressing difficulty 3).

We illustrate how the instabilities arise due to the Kozai-Lidov (KL) mechanism
in wide ($\gtrsim100$ AU) stellar binaries, although our proposal 
is more general and sub-stellar companions and other sources of
secular excitation are allowed.
We expect that for these wide binaries the possible WD pollution 
associated with post-AGB dust disks 
and stellar winds  might be negligible
(e.g., \citealt{ruyter06,VW09,bilik12,clayton14}).

\section{Planet Engulfment as a Trigger for ``Dormant" Secular Instabilities}
\subsection{Planetary Systems as Suppressors of Secular Instabilities}
White dwarf pollution by tidally disrupted minor rocky bodies requires
a mechanism to deliver asteroids from distant orbital separations 
into the stars's tidal disruption radius ($\sim1R_\odot$). Nearly radial 
orbits may result
from secular instabilities, which in some cases are capable
of exciting eccentricities
up to values of $\sim1$.
One well-known example of such instabilities is 
the KL mechanism (\citealt{kozai,lidov};  
see \citealt{naoz16} for a recent review), which takes place
when a distant stellar-mass companion is highly inclined respect to
the orbit of the minor body. However, it is also known that
additional bodies in the system may affect or entirely
suppress the effect of the KL mechanism
\citep[e.g.][]{holman97}. 

In the simplest scenario
of one planet in a circular orbit with mass $M_{\rm p}$ and semi-major axis
$a_{\rm p}$ inside a planetesimal's orbit ($a_{\rm p}<a$), 
the effect of the additional quadrupole potential due to the planet's
time-averaged orbit will overcome that of the outer stellar
companion if the planet-induced apsidal precession frequency 
\ba
\dot{\varpi}_\mathrm{in}
\simeq
\frac{1}{2} n \left(\frac{M_{\rm p}}{M_{\rm s}}\right)\left(\frac{a_{\rm p}}{a}\right)^2
\approx n\epsilon\in,
\label{eq:v_in}
\ea
is larger than that induced by a binary
with mass $M_{\rm b}$ and semi-major axis
$a_{\rm b}$ of
\ba
\label{eq:omega_out}
\dot{\varpi}_\mathrm{out}
\simeq
n \left(\frac{M_{\rm b}}{M_{\rm s}}\right)\,\left(\frac{a}{a_{\rm b}}\right)^3(1-e_{\rm b}^2)^{-3/2}
= n\epsilon\out,
\label{eq:varpi_out}
\ea
with $n$ being the mean motion frequency of the planetesimal and
where we have used the definition of two dimensionless quantities $\epsilon\in$ and
$\epsilon\out$ that represent the relative strength of the tidal potentials (see the Appendix and \citealt{mun15b}).

When $\dot{\varpi}_\mathrm{in}=\dot{\varpi}_\mathrm{out}$, then $a=r_L$, 
where $r_L$ is the ``Laplace radius'', defined as
\ba
r_{\rm L}&\equiv&\left(\frac{M_{\rm p}}{2M_{\rm b}}a_{\rm p}^2
a_{\rm b}^3\left[1-e_{\rm b}^2\right]^{3/2}
\right)^{1/5}\nonumber\\
&\simeq&16.2 ~\mbox{AU}~\left(\frac{M_{\rm p}}{M_J}\right)^{1/5}
\left(\frac{M_{\rm b}}{0.5M_\odot}\right)^{-1/5}
\left(\frac{a_p}{2~\mbox{AU}}\right)^{2/5}\nonumber\\
&&\times\left(\frac{a_{\rm b}\sqrt{1-e_{\rm b}^2}}
{600\sqrt{1-0.5^2}~\mbox{AU}}\right)^{3/5}.
\label{eq:rl}
\ea

For $a<r_L$, the dynamics of the asteroid will be dominated by
the planet's quadrupole potential, such that the planetesimal's 
angular momentum vector $\propto\mathbf{j}$
will precess around the planet's, with perfect alignment being the
equilibrium solution. Conversely, for $a>r_L$, the dynamics of the asteroid
will be dominated by the binary companion, with $\mathbf{j}$
 precessing around the binary's angular momentum vector
(with the possibility of being Kozai-unstable), with
perfect alignment being the equilibrium solution. The smooth 
transition between these
two regimes place takes rapidly around $a\simeq r_L$, and the
 general equilibrium solution of the equilibrium inclination $i_{\rm eq}$
 for all values of  $a$ is known
 as the ``Laplace surface".  For a test particle in a 
 circular orbit, the Laplace surface
is given by (e.g., \citealt{tremaine09,tamayo13}):
\ba
\tan 2 i_{\rm eq}=\frac{\sin 2i_{\rm b}}{\cos 2 i_{\rm b}+2(r_{\rm L}/a)^5}
\label{eq:i_lap}
\ea
where $i$ ($i_{\rm b}$) is the inclination 
of the test particle (binary)
relative to the planetary system

 Thus, for as long as there is a planet (or a planetary system)
 such that $r_L$ is large enough to accommodate a (nearly)
 coplanar population of planetesimals/asteriods, 
 such bodies will be protected from the tidal 
 potential from the binary companion,
 largely ignoring its presence throughout the main sequence 
 (MS) evolution of the host star.

\subsection{Triggering of secular instabilities}
\label{sec:trigger}
Any reduction of the quadrupole potential due to the planet
will reduce the extent of the ``safe zone" defined by the Laplace radius,
progressively exposing bodies to the influence of the binary companion's
tidal potential. One possible cause of such a change is planetary engulfment during the
post main sequence stages of stellar evolution. During the red giant branch
(RGB) and asymptotic giant branch (AGB) phases
 of the post-MS, low-to-intermediate
mass stars can reach radii of a fraction of, or up to few AU, presumably engulfing
all planets within this distance \citep{MV12,villa14}. 
In particular, the AGB phase
is during which most of the mass in the stellar envelope is lost, resulting in the 
expansion of all the orbits in the system.

The engulfment of a planet consists of both its evaporation and its spiraling in
during stellar expansion (potentially aided by the tidal interaction with the extended
stellar envelope; \citealp{villa14}). For simplicity, here we simply model the engulfment
as the gradual reduction of the planetary semi-major axis $a_p$:
\begin{equation}
a_p(t)=a_{p,0}e^{-t/\tau_a}\;\;\; \text{  for  }\; t_{\rm MS}<t<t_{\rm WD}
\label{eq:aplanet}
\end{equation}
where $\tau_a$ represents the in-spiral timescale of the planet, $t_{\rm MS}$ is
the duration of the stellar MS and $t_{\rm WD}$ is the time at which the WD
is formed. 
As $r_L\propto a_{\rm p}^{2/5}$ (Eq.~\ref{eq:rl}), planetary engulfment causes
the Laplace radius the decrease.

In addition, we consider mass loss. If the expulsion of the stellar outer layers
happens on timescales much longer than all the orbital periods in the system,
then angular momentum conservation dictates that all semi-major axes evolve
as $a/\dot{a}=-M_{\rm s}/\dot{M_{\rm s}}$  \citep[e.g.,][]{hadjidemetriou63,veras11}.
Thus, introducing another timescale
$\tau_{\rm ml}$ we can write:
\begin{equation}
M_{\rm s}(t)=\left\{
\begin{array}{lc}
M_{\rm s,0}e^{-t/\tau_{\rm ml}}\;\text{  if  }& M_{\rm s,0}e^{-t/\tau_{\rm ml}}> M_{\rm WD}\\
M_{\rm WD} & \sim
\end{array}
\right.
\label{eq:massloss}
\end{equation}
for  $t>t_{\rm MS}$.
Which in turn implies $a(t)=a(0)e^{t/\tau_{\rm ml}}$ and 
$a_b(t)=a_{b}(0)(M_{{\rm s},0}+M_{\rm b})/(M_{{\rm s},0}e^{-t/\tau_{\rm ml}}+M_{\rm b})$
for $t_{\rm MS}<t<\tau_{\rm ml}\ln(M_{{\rm s},0}/M_{\rm WD})$.

The effect of mass loss has an opposite effect to engulfment on the value of the Laplace radius
($r_L\propto a_b^{3/5}$).
For $M_b\ll M_s$, one can write an approximate evolution of the Laplace radius:
\ba
 \frac{r_{\rm L}(t)}{r_{\rm L}(0)}=   
 \exp{\Big[-\cfrac{2t}{5\tau_{\rm a}}+\cfrac{3t}{5\tau_{\rm ml}}\Big]},
~t_{\rm MS}<t<\tau_{\rm ml}\ln\left(\tfrac{M_{{\rm s},0}}{M_{\rm WD}}\right)\nonumber\\
\ea
which shows how mass loss and engulfment have opposite effects. 
Ultimately, engulfment is the determining factor, since $r_L,\rightarrow0$ as $a_{\rm p}\rightarrow0$, 
while mass loss stops when the stellar remnant mass reaches $M_{\rm WD}$\footnote{Note that
a planet is engulfed when it reaches the stellar envelope at $a_{\rm p}\simeq R_s$ at some point
of the evolution (e.g., \citealt{MV12,villa14}), which 
is equivalent to setting $a_{\rm p}=0$ in $r_{\rm L}$ because the quadrupole moment
provided by the planet vanishes after engulfment (i.e., $\dot{\varpi}_{\rm in}=0$ in Equation
[\ref{eq:v_in}]).}.

The reduction of $r_L$ has a direct impact on the shape of the Laplace surface
in Eq.~(\ref{eq:i_lap}). 
(1) If  the ``parameter" $r_L$ changes {\it very slowly}, the $i_{\rm eq}$ curve
will evolve smoothly, and any bodies initially lying on the Laplace surface will remain
on the final Laplace surface\footnote{Note that a dynamical solution can ``follow'' a slowly changing fixed point 
{\it provided} this equilibrium remains stable. If the stability of the fixed point changes
-- i.e., it experiences a bifurcation -- then adiabaticity is broken. The Laplace surface is known to become
unstable in eccentricity (circular orbits are not allowed) 
in a narrow region of $a$ around $r_L$ only if $i_{\rm b}\gtrsim 69^\circ$ \citep{tremaine09}
}. 
This implies full coplanarity with
the binary companion when 
$r_L\rightarrow0$, i.e., $i_{\rm final}\sim i_{\rm eq}\rightarrow i_{\rm b}$. 
(2) On the contrary,
if $r_L$ is changed {\it very rapidly},
then the  
bodies initially lying on the Laplace surface will not be able to catch up, thus being
frozen in their initial inclinations $i_{\rm final}\sim i_{\rm initial}\sim0$. These two limits
entail completely opposite consequences for the planetesimals: (1) if the planet is removed slowly,
planetesimals will reach coplanarity with the companion, thus being Kozai-stable; (2) if the planet
is removed suddenly, minor bodies will see their initial inclinations unchanged, 
thus being
susceptible to KL oscillations. 

\subsubsection{Adiabaticity}\label{sec:adiabaticity}
%
The speed at which $r_L$ is changed  (either ``very slow'' or ``very fast'' as defined below)
will determine whether
configurations near equilibrium evolve ``adiabatically" or not \citep[e.g.][]{landau}.
Quantitatively,
the rate of change in $\epsilon\in$ (or in $r_L$)
must be much smaller than the linear oscillation frequency $\omega_0$
around the equilibrium solution of the Laplace surface. Rewriting eq. (31) in \citet{tremaine09}
as
\begin{equation}\label{eq:osc_freq}
\begin{split}
\omega_0&=\frac{3}{2\sqrt{2}}n\epsilon\in\Bigg[
1+\cos2i_{\rm eq}-\sin2i_{\rm eq} \\
&~~~~~~~~~+\frac{1}{2}\left(\frac{\epsilon\out}{\epsilon\in}\right)
\left(\cos2i_{\rm eq}+2\cos2i_{\rm b}\right)\\
&~~~~~~~~~+\frac{1}{2}\left(\frac{\epsilon\out}{\epsilon\in}\right)^2\cos^2i_{\rm b}
\Bigg]^{1/2}~~.
\end{split}
\end{equation}
The degree of adiabaticity can be represented by the ratio $|(\dot{r_L}/r_L)|/\omega_0$
\citep{landau}, which
is roughly 
\begin{equation}
\frac{2}{3}\frac{\tfrac{2}{5}\tau_a^{-1}}{n\epsilon\in}
\sim \frac{1}{\dot{\varpi}\in\tau_a}
\;\;\text{if}\;\;\;\epsilon\out\ll\epsilon\in~~,
\end{equation}
or
\begin{equation}
\frac{4}{3}\frac{\tfrac{2}{5}\tau_a^{-1}}{n\epsilon\out\cos i_{\rm b}}\sim\frac{1}{\dot{\varpi}\out \tau_a\cos i_{\rm b}}\sim \frac{\tau_{\rm KL}}{\tau_{\rm a}\cos i_{\rm b}}
\;\;\text{if}\;\;\;\epsilon\in\ll\epsilon\out~~,
\end{equation}
where we have introduced the KL timescale \citep[e.g.,][]{anto15}
\begin{equation}\label{eq:tau}
\begin{split}
\tau_{\rm  KL}&=\frac{16}{15\pi}\left(\frac{a_{\rm b}
\sqrt{1-e_{\rm b}^2}}{a}\right)^{3}
\frac{M_{\rm s}}{M_{\rm b}}P\\
&\simeq2.3~ \mbox{Myr} 
\bigg(\frac{a}{10\mbox{ AU}}\bigg)^{-\tfrac{3}{2}}
 \bigg(\frac{a_{\rm b}}{600\mbox{ AU}}\bigg)^{3}({1-e_{\rm b}^2})^{3/2}\\
 &~~~\times\bigg(\frac{M_{\rm s}}{M_\odot}\bigg)^{\tfrac{1}{2}} \bigg(\frac{M_{\rm b}}{M_\odot}\bigg)^{-1}.
\end{split}
\end{equation}
Thus, after $r_L$ has shrunk below $a$, the two relevant timescales to
compare will be $\tau_{\rm a}$ and $\tau_{\rm KL}$, and for evolution
on the Laplace surface to be adiabatic, it is required that
$\tau_{\rm a}\gg\tau_{\rm KL}$. As we will see below, this
requirement will be rarely met by planet engulfment, and thus adiabaticity
will be most commonly broken as $r_L\rightarrow0$.
For this reason, in most situations, planetesimals will be suddenly exposed 
to the KL mechanism once planet engulfment has been completed (see Fig.~\ref{fig:phases} for a schematic depiction).  
Thus, only
after engulfment has removed the
protection against
eccentricity excitation, will the external perturber be
enabled to place planetesimals into 
orbits leading to their disruption.


\subsection{Toy Model of Four Bodies with Octupole Terms}
In the absence of a planet, planetesimals will be susceptible to KL oscillations induced by
the stellar binary provided that the relative inclination between the planetesimal and the 
binary $i_{\rm b}$ is in the range of $40^\circ-140^\circ$. 
Nevertheless, for planetesimals with $a\sim10$~AU to
reach the tidal disruption distance of $\sim R_\odot$ at pericenter, their eccentricities
must be such that $a(1-e)\sim R_\odot=5\times10^{-3}$~AU, i.e $(1-e)\sim 10^{-3}-10^{-4}$,
which is difficult to achieve during quadrupole-order KL oscillations. 
The eccentric KL mechanism (\citealp{naoz16}, which includes
 octupole-order terms in the tidal potential when $e_{\rm b}\neq0$), 
 on the other hand, is capable of producing such high eccentricities,
thus satisfying the requirements for tidal disruption of planetesimals.

In order to study the evolution of planetesimals during the stellar MS and through the post-MS 
we integrated the (double-averaged) secular equations of a hierarchical four-body
system \citep{ham15,mun15b} consisting of
 a star (of mass $M_{\rm s}$), a gas giant
(of mass $M_{\rm p}$ and semi-major axis $a_{\rm p}$),
a massless particle (semi-major axis $a$) and a distant companion
(of mass $M_{\rm b}$ and semi-major axis $a_{\rm b}$). The initial setup is reminiscent
of that of \citet{mun15b} but in this case we have included
octupole-level terms in the potentials following \citet{ham15} (see the Appendix
of the current manuscript).

\subsubsection{Equations of Motion}

We evolve the planetesimal's
dimensionless specific angular momentum vector
$\jvec=\sqrt{1-e^2}~\hat{\bf{j}}$
and eccentricity vector $\evec$
according to the equations of motion \citep[e.g.][]{tremaine14}:
\begin{subequations}\label{eq:eom}
\begin{align}
\label{eq:motion_tp_j}
\frac{d\jvec}{dt}=\frac{1}{\sqrt{\mathcal{G}M\in a}}\Big(
\jvec\times\nabla_\jvec \Phi
+\evec\times\nabla_\evec \Phi
\Big),\\
\label{eq:motion_tp_e}
\frac{d\evec}{dt}=\frac{1}{\sqrt{\mathcal{G}M\in a}}\Big(
\jvec\times\nabla_\evec \Phi
+\evec\times\nabla_\jvec \Phi
\Big),\
\end{align}
\end{subequations}
where $M\in=M_{\rm s}+M_{\rm p}$ and 
$\Phi=\Phi\in+\Phi\out$ is the combined tidal potential from the inner star-planet pair and the outer
stellar companion including octupole-order terms 
(see Eqs~[\ref{eq:potential_av}]-[\ref{eq:potential_av_gr}]).
 The star-planet angular momentum and eccentricity
vectors $\propto\jvec\in$ and $\evec\in$  are evolved following an analogous
set of equation of motion, except that the tidal potential is entirely due to
the binary companion, with the contribution of the planetesimal being neglected.
Finally, the angular momentum and eccentricity vectors of the star-binary system 
$\propto\jvec\out$ and $\evec\out$ -- although in principle subject to evolution under
the tidal potential from the inner star-planet pair -- are held constant and only
the semimajor axis of this outermost orbit $a\out$ is evolved consistently with
the process of mass loss of the central star.

\begin{figure*}
   \centering
  \includegraphics[width=16cm]{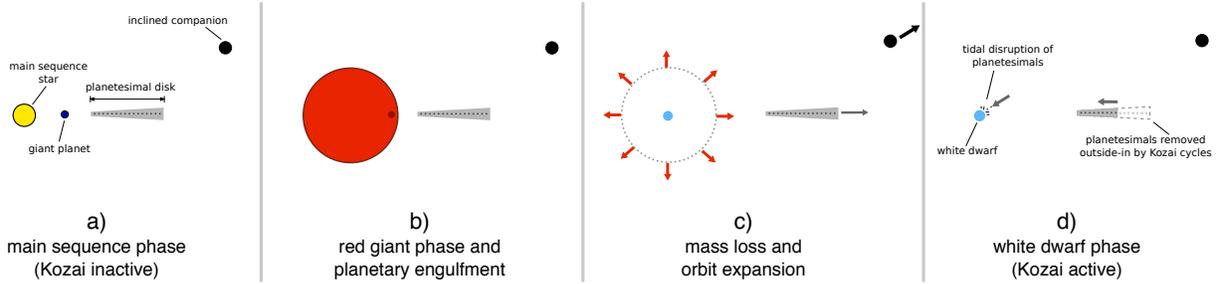}
  \caption{The orbital architectures at the different phases of the stellar evolution considered in our example. {\it Panel a}: a $2M_{\odot}$ MS star orbited by a giant planet at $a_{\rm p}=2$ AU, a coplanar planetesimal disk at $a=3-12$ AU, and inclined binary companion ($i_{\rm b}=80^\circ$) at $a_{\rm b}=600$ AU. The disk remains stable against the KL mechanism because of the planet perturbations. 
  {\it Panel b}: the planet is engulfed by the host star during the giant phase (GB or AGB phases).
  {\it Panel c}: the orbits of the planetesimal disk and the binary expand due to mass loss.
  {\it Panel d}: the disk is subject to the KL mechanism and the planetesimals are tidally disrupted outside-in.
 }
\label{fig:phases}
\end{figure*}

\subsubsection{Initial set-up: main-sequence configuration}
\label{sec:setup}
We consider a host star with a Zero Age Main Sequence mass
of $M_{\rm s}=2M_\odot$, which is a typical progenitor
for the currently-observed WD population in the Milky Way.
We assume that this star is orbited by a Jupiter-mass planet 
at $a_{\rm p}=2$ AU  in a circular orbit, 
and a wide stellar binary companion with a mass of $M_{\rm b}=0.5M_\odot$,
a semi-major axis of $a_{\rm b}=600 $ AU and an 
eccentricity of $e_{\rm b}=0.5$ (see panel a in Figure \ref{fig:phases}). The
inclination relative to the planetary orbit 
is $i_{\rm b}=80^\circ$.
The planet is subject to the tidal field from the stellar companion
and apsidal precession due to General Relativity (GR). At 2~AU, 
the apsidal precession period due to GR  ($\sim10$ Myr) is shorter
than that due to the companion ($\sim30$~Myr; Eq. [\ref{eq:tau}]),
and thus  KL oscillations are suppressed. In practice, other sources of pericenter 
precession such as additional planets 
 can make the planetary orbit long-term stable against perturbations
 from the inclined companion. In this configuration, $r_L\approx16$~AU,
 and thus any bodies interior to this distance will  be protected by the planet
 against perturbations from the binary companion.
The fourth body in the system
is a massless planetesimal located at $a=10$~AU -- such that
$a<r_L$ -- in near coplanarity with the planetary orbit 
($i_{\rm eq}\simeq0.46^\circ$ from Eq. [\ref{eq:i_lap}])

\subsubsection{Post-main sequence evolution}
The system is evolved at once  (including MS and Post-MS stages, 
see the sequence depicted in Figure \ref{fig:phases})
under one set of equations (Eq.~[\ref{eq:eom}]).
After some time $t_{\rm MS}$, mass loss and planetary engulfment are triggered
(Eqs.~[\ref{eq:aplanet}] and [\ref{eq:massloss}]), affecting
directly the semi-major axes $a_{\rm p}$, $a_{\rm b}$ and $a$ (which are constant in
the secular evolution of the MS stage), in addition to the central mass $M_s$
(see Eqs.~(\ref{eq:ml_in})-(\ref{eq:ml_out}) in the Appendix).  The timescale
for planetary removal and mass loss is expected to be of a few Myr, and to take place
primarily during the AGB phase of stellar evolution (e.g., \citealt{hurley00}).
 We set $\tau_{\rm ml}=4\tau_a$ to ensure
 that the mass loss and the orbit's shrinkage happen
simultaneously, but the former takes place mostly at the end
of the planet's orbital decay.  
During this phase, planets inside $\sim3-5$ AU are expected to be 
engulfed by the expanded
envelope of the host star \citep{MV12,villa14}. By using the secular equations
of motion it is implicitly assumed that the mass loss and engulfment timescales
are much longer than all the orbital timescales in the system. For outer companion
separations of up to $\sim2000$~AU, this is a reasonable approximation.  In such
case, all the orbital elements in the system, except the semi-major axes,
will remain unchanged \citep[e.g.,][]{hadjidemetriou63,veras11}. 

In our example, we use that the final mass -- the mass of the WD -- is 
$M_{\rm WD}\simeq0.64M_\odot$ \citep{hurley00}.
The semi-major axes of 
the small bodies expands in factor of 
$M_{\rm s}(t=0)/M_{\rm WD}\simeq3.1$,
while the binary does so in a factor
of $[M_{\rm s}(t=0)+M_{\rm b}]/(M_{\rm WD}+M_{\rm b})\simeq2.2$.
Initially, $\epsilon\in/\epsilon\out=(r_{L,0}/a)^5\approx11$. As the planet
is engulfed, $\epsilon\in/\epsilon\out\rightarrow0$; in practice, the potential
from the planet is ignored after $\epsilon\in/\epsilon\out$ reaches
$10^{-5}$, or when $r_L=0.1 a$.

After the planet influence becomes negligible, the planetesimals
pericenter precession will be given by $\dot{\varpi}\out$ (Eq.~\ref{eq:omega_out}).
If the removal of the planet is non-adiabatic (as defined in Section~\ref{sec:adiabaticity}),
the planetesimal may be subject to KL oscillations, which take
place with a characteristic period of $\tau_{\rm KL}\sim \dot{\varpi}\out^{-1}$. In classic
Kozai oscillations (quadrupole-order perturbations  $\propto a^2/a_{\rm b}^3$) 
the planetesimals eccentricity may reach
$e_{\rm max}=(1-5/3\cos^2 i_{\rm b})^{1/2}\simeq0.97$,
implying a minimum pericenter
distance of $a(1-e_{\rm max})\simeq1$ AU. However, for $e_{\rm b}\neq0$,
the planetesimal 
is subject to strong forcing due to higher-order 
(mostly octupole-order $\propto e_{\rm b}a^3/a_{\rm b}^4$) 
perturbations, which happen in
timescales longer than the KL timescale \citep[e.g.][]{naoz16}.
These {\it very long timescale} effects can drive the eccentricity up to  much higher values
 ($1-e\lesssim0.001$). If high enough eccentricities are reached such that $a(1-e)<2R_\odot$,
then the planetesimal is assumed to be tidally disrupted.

\begin{figure*}
   \centering
  \includegraphics[width=17cm]{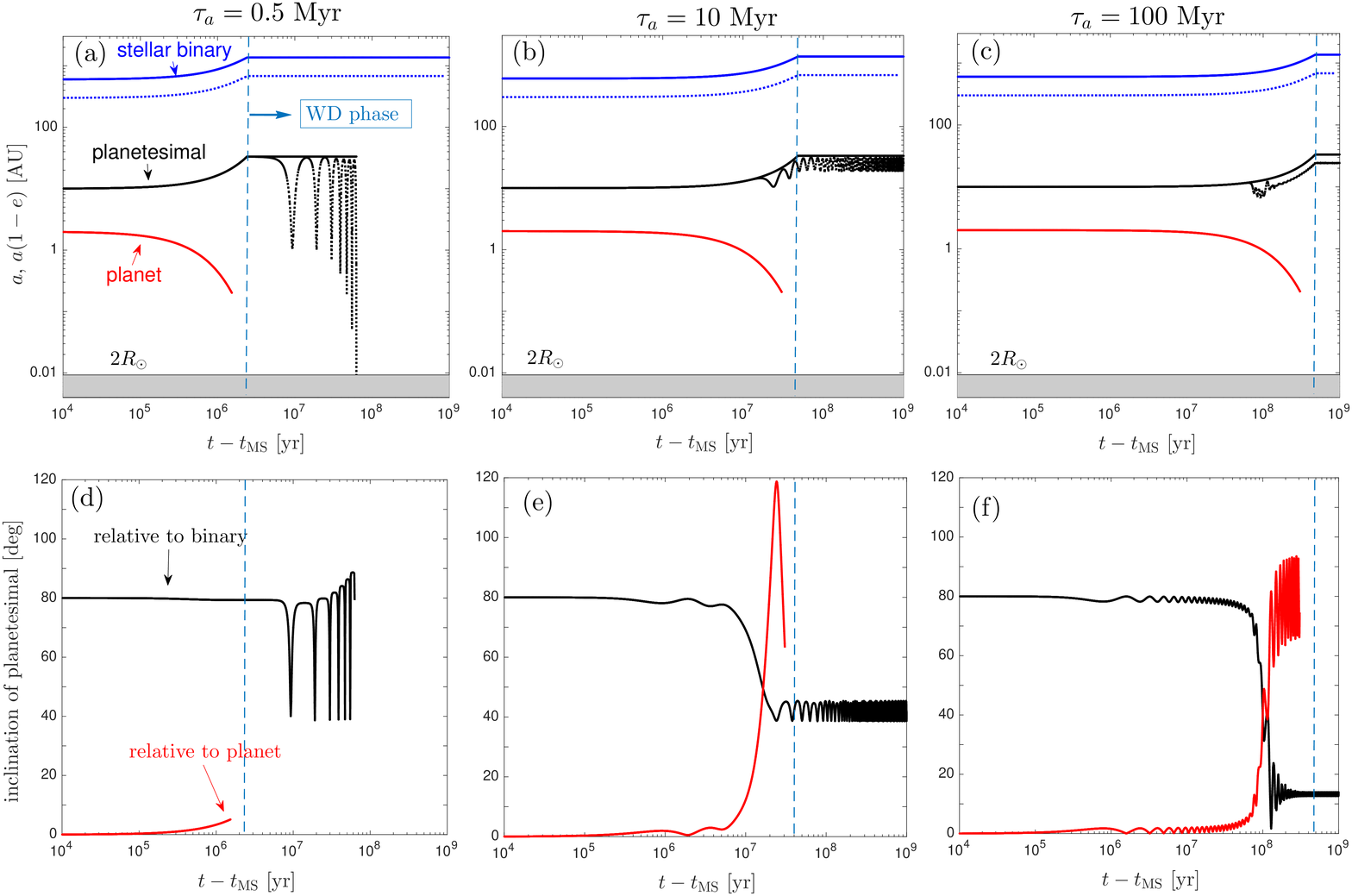}
  \caption{Orbital evolution of the planetary system 
  initially composed of an inner Jupiter-mass planet at 2 AU and a massless planetesimal  at 10 AU orbiting  a 2$M_\odot$ star (final WD mass of 0.64 $M_\odot$). 
  We show the results after the main sequence ($t>t_{\rm MS}$)
  and the vertical dashed lines indicate the zero WD's cooling age.
  The planetesimal and the planets start
 with zero mutual inclination and circular orbits, while the binary companion with $M_b=0.5M_\odot$
 and $a_b=600$ AU
 has an inclination of $80^\circ$ relative to the planetary system. The eccentricity
 of the binary is $e_b=0.5$, and the initial ascending nodes and arguments of pericenter
 are 0 for all the orbits.
The upper panels (a, b, and c) show the semi-major axes (solid lines) and 
pericenter distances (dotted lines).
The lower panels (d, e, and f) show the inclination of the planetesimal relative to the
planets (solid red line) and relative to the binary (solid black line).
The different columns show the different planet's semi-major axes 
decaying timescale $\tau_a$ in Equation (\ref{eq:aplanet}): 
$\tau_a=0.5$ Myr (panels a and d),
$\tau_a=10$ Myr (panels b and e),
and $\tau_a=100$ Myr (panels c and f).
The planet is assumed to be engulfed at 0.2 AU and the planetesimal is 
assumed to be tidally disrupted
when it reaches $a(1-e)<2R_\odot$, which only happens for 
$\tau_a=0.5$ Myr (dotted black line in panel a).
We set the mass loss timescale $\tau_{\rm ml}=M_s/|\dot{M}_s|$ equal
to $4\tau_a$ to ensure that mass is lost most efficiently 
after the planet is engulfed.
 }
\label{fig:tau_a}
\end{figure*}

\subsubsection{Fast and slow planet engulfment}
To directly test the qualitative predictions of Section~\ref{sec:trigger},
we integrate the 4-body equations of motion while varying 
the engulfment time $\tau_{\rm a}$. As described in Section~\ref{sec:adiabaticity},
the ratio $\tau_{\rm a}/\tau_{\rm KL}$ will determine whether or not
the planetesimal will be susceptible to the influence of the eccentric KL mechanism.

\paragraph{Fast engulfment ($\tau_{\rm a}=0.5$~Myr)} For $\tau_{\rm a}\ll \tau_{\rm KL}\sim 8$~Myr
(left panels, Fig.~\ref{fig:tau_a}), we expect adiabaticity to be
broken and the planetesimal to be impulsively removed from
the Laplace surface. In this case, orbits expand according to
the mass loss prescription (top panel), but this takes place
before the planetesimal inclination has been altered significantly (bottom panel).
The planetesimal-to-binary inclination is this
 the same as in the initial condition ($\sim80^\circ$), which is capable of triggering KL
oscillations. After planet removal, common KL oscillations ensue (with period
of $\simeq10$~Myr), reaching a maximum planetesimal eccentricity of 0.9.  
 In the longer run, 
the slower octupole-level oscillations cause  dramatic eccentricity growth,
reaching $e\simeq0.999$ and beyond, sufficient to guarantee tidal disruption.

\paragraph{Trans-adiabatic engulfment ($\tau_{\rm a}=10$~Myr)}
When $\tau_{\rm a}\sim\tau_{\rm KL}$ (Fig.~\ref{fig:tau_a}, middle
panels), the orbital behavior of the planet is markedly different
from the ``fast engulfment'' case from above.
In this case, the planetesimal ``tries to follow'' the
Laplace surface solution $i_{\rm eq}(r_L)$ (Eq.~[\ref{eq:i_lap}]) 
as $r_L$ shrinks with time. Initially, the planetesimal
can follow closely; the initial oscillation amplitude
($\lesssim2^\circ$) is seeded by the imperfect alignment
of $i_0$ with $i_{\rm eq}$ at $t=0$. If adiabaticity were to be preserved,
this initial amplitude should grow as the oscillation frequency decreases
from $\sim\dot{\varpi}\in|_{t{=}0}$ to $\sim\dot{\varpi}\out$
\footnote{Linear oscillations around the stable equilibrium should behave like a harmonic oscillator
of time-varying frequency $\omega_0(t)$ (Eq.~\ref{eq:osc_freq}), for which the action is $E/\omega_0=\omega_0A^2/2$,
where $A$ is the oscillation amplitude. For very slowly varying $\omega_0$, the action
$E/\omega_0$ is an adiabatic invariant.}. 
Nevertheless, planet engulfment is still too fast, as the planetesimal leaves the Laplace surface
before the planet is fully engulfed. In this case, the planetesimal is decoupled from the planet at 
$t{\simeq}20$~Myr (the planet is finally eliminated at $t{\simeq}30$~Myr), 
with an inclination relative to the binary of $\gtrsim40^\circ$. This small inclination ($>39.23^\circ$)
is still large enough to trigger some mild KL oscillations, but evidently it is far from the inclinations needed
to obtain tidal disruptions as $e_{\rm max}$ is only $\simeq0.45$.

\paragraph{Slow engulfment ($\tau_{\rm a}=100$~Myr)}
For slow engulfment ($\tau_{\rm a}\gg\tau_{\rm KL}$, right panels, Fig.~\ref{fig:tau_a}), 
the planetesimal nearly follows the Laplace surface to the end of the
integration, reaching a final inclination of only $\sim13^\circ$ and a finite constant eccentricity
of  $\simeq0.3$. The finite eccentricity  is reached at $\simeq60$~Myr when $a\sim r_L$, and is
due to the bifurcation experienced in the Laplace equilibrium solution, which makes circular orbits unstable
\citep{tremaine09,tamayo13}. For even slower engulfment ($\tau_{\rm a}=1$~Gyr), the final state is much more
steady, i.e., inclination oscillations are small, and the planetesimal never leaves the vicinity of the Laplace surface, even if it gains a finite eccentricity
as it crosses the bifurcation. We note that this bifurcation exists only when the planet-binary inclination is $\gtrsim69^\circ$.
In addition, the finite octupole potential from the binary introduces modifications to the classical Laplace equilibrium
analysis \citep[e.g.][]{mun15b}. In an analogous example with $e_{\rm b}=0$ and a lower inclination
we obtain an end-state where the planetesimal is in perfect alignment with the binary and retains zero eccentricity throughout 
the integration.

\vspace{5mm}
Of these three scenarios,
only the first one (fast engulfment) is expected to
resemble the AGB phase, which lasts less than $\sim10$ Myr.
This introduces an important difference with the work of \citet{mun15b},
which finds that the rate of reduction of $r_L$ is always slower than
the oscillations around the equilibrium solution.  In the present case,
adiabaticity is an unlikely outcome, which implies that planetesimals
can ``instantaneously" see themselves in a Kozai-unstable 
configurations even though
throughout the entire MS lifetime of the host star they were protected
against such instabilities. 

Finally, we caution that our numerical calculations
are based  on the double orbit
averaging approximation and might not represent the 
dynamics properly, leading to spurious extreme eccentricities 
required  to disrupt the planetesimal 
\citep{LKD16}. 
To this extent, we have repeated the three-body
integrations after planet engulfment using
the direct high-order N-body
 integrator IAS15 \citep{RS15}, 
which is part of the REBOUND package \citep{RL12}.
We find that the evolution looks very similar, but the
planetesimal is disrupted slightly later after two extra
oscillation cycles compared with the secular code.


\section{Evolution of the planetesimal disk}

We integrate the orbital evolution of a disk of collision-less
planetesimals  orbiting a white dwarf and perturbed by
a distant companion based on the orbital configuration of
our fiducial system described in \S\ref{sec:trigger}.
This phase corresponds to panel d in Figure \ref{fig:phases}.

\subsection{Initial conditions}

After the AGB phase we are left with a 
a WD of mass $M_{\rm WD}\simeq0.64M_\odot$ 
orbited by a planetesimal disk and stellar binary
companion with mass of $M_{\rm b}=0.5M_\odot$,
semi-major axis $a_{\rm b}=1300$ AU, and
inclination of $i_{\rm b}=80^\circ$ (relative to the planetary 
orbit before engulfment).

We shall assume that the planetesimal disk 
has the following power-law profile
for the surface density
\ba
\Sigma(a)=\frac{M_{\rm disk}}
{2\pi\left(a_{\rm out}-a_{\rm in}\right)}\cdot\frac{1}{a},
\ea
where $M_{\rm disk}$ is the total mass of the disk, 
while $a_{\rm in}$ and $a_{\rm out}$
are its inner and outer boundaries. 
This power-law profile has a uniform mass distribution
as a function of semi-major axis, which is a convenient 
choice to easily read out our results for the accretion
rates.

The inner edge of the disk $a_{\rm in}$ is set by the dynamical stability
due to the planetary perturbations during the Main Sequence.
Since the planet is initially at $a_{\rm p}=2$ AU, the long-term stability
is guaranteed for planetesimals at $\gtrsim3$ AU\footnote{The 
test particles are Hill stable for $a/a_p\gtrsim1.4$
(e.g., \citealt{gladman93}).}.
Since the semi-major axis of the planetesimals
expands by a factor of $\simeq3.1$ relative to the initial value
during the MS, we set $a_{\rm in}=10$ AU.

The outer edge is set by the location of the Laplace radius, beyond
which the secular perturbations due to the binary can 
excite the eccentricities and inclinations of the planetesimals.
More specifically, as shown by \citet{tremaine09} 
for a binary's inclination of $i_{\rm b}=80^\circ$ a test particle
is stable against eccentricity perturbations when
$a<0.9r_{\rm L}\simeq14.6$ AU. 
Thus, we conservatively consider the planetesimals with
$a<12$ AU in our calculations so the disk remains nearly
aligned with the orbit of the planet
and with low-eccentricity excitation during the planet 
engulfment.
For reference, the Laplace surface  of
a planetesimal with $a=12$ AU
has an inclination relative to the 
planetary orbit of $i_{\rm eq}\sim1.1^\circ$ (Eqs. [\ref{eq:i_lap}] 
and [\ref{eq:rl}]).
Again, since the semi-major axis of the planetesimals
expands by a factor of $\simeq3.1$ relative to the initial value
during the MS, we set $a_{\rm out}=35$ AU.

The planetesimal orbits are initialized with zero inclination, 
random longitudes of the ascending 
node and arguments of pericenter, and 
eccentricities from a Rayleigh distribution
with parameter 0.01.

We leave the mass of the disk as a free parameter, but
 keep in mind that the collisional evolution of the disk
 limits the maximum mass for a given age 
 (e.g., \citealt{wyatt07,HT10}).
We discuss the constraints on the disk mass in 
\S\ref{sec:discussion}.

\begin{figure*}
   \centering
  \includegraphics[width=18cm]{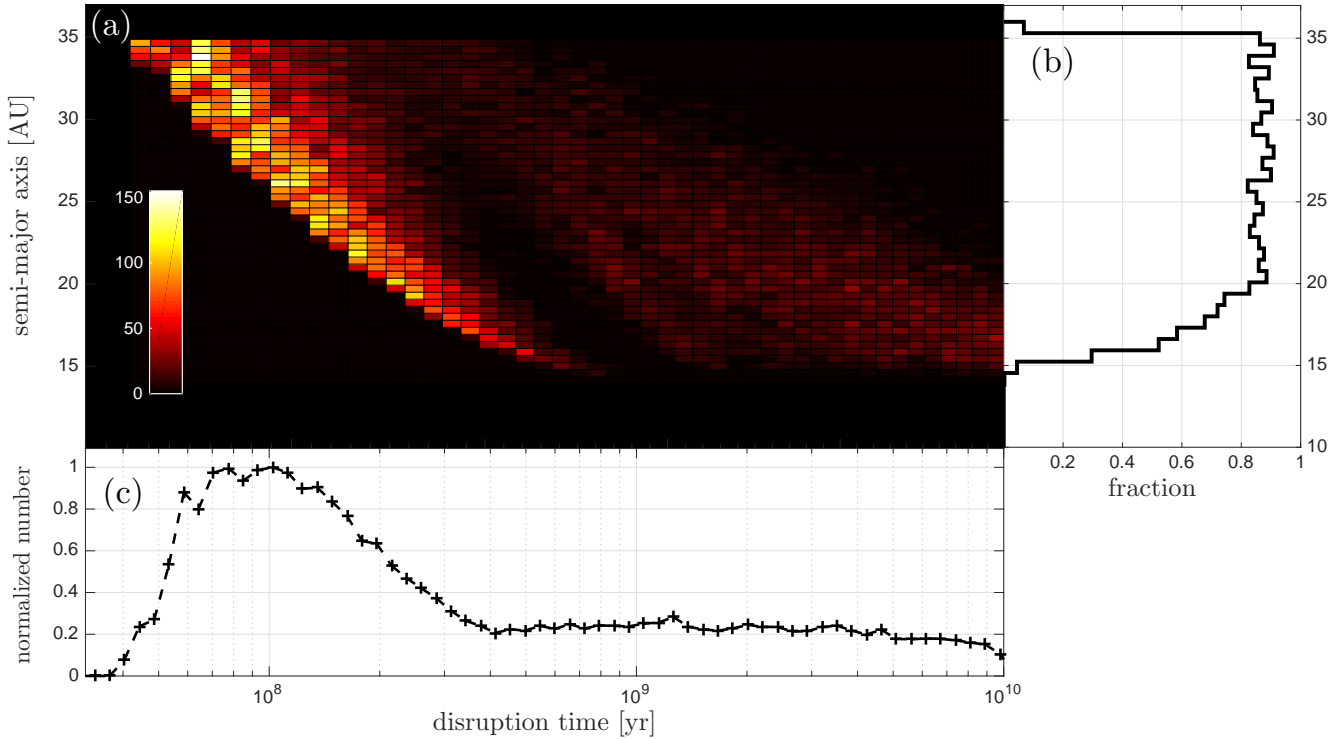}
  \caption{Number of particles that are tidally disrupted
  by the WD  as a function of their initial semi-major axis during the WD phase (i.e., after their orbits expanded by a factor of $\simeq3.1$ relative to their MS values) and
  the time at which they cross the disruption distance 
  (initial time is the zero cooling age time).
  {\it Panel a:} two-dimensional histogram. The binning in the horizontal axis is evenly-spaced in the log of the disruption time.
  {\it Panel b:} fraction of tidally disrupted particles
  as a function of semi-major axis.
  {\it Panel c:} number of tidally disrupted particles normalized 
  by the tallest bin as a
  function of the disruption time.
 }
\label{fig:pop_dis}
\end{figure*}

\subsection{Results}

We evolve the disk up to 10 Gyr 
using $50,000$  particles
and record the time at which a particle is tidally disrupted,
which we define to takes place once
$a(1-e)<2R_\odot$.
The results are not sensitive to the choice of the disruption
distance, and typical values within $\sim1-3R_\odot$ 
\citep{veras14b} give similar results.

In Figure \ref{fig:pop_dis} we show the number 
of disrupted bodies as function of the initial
semi-major axes and the disruption times.
From panel a we observe that  the bodies 
at larger semi-major axes tend to be
disrupted first.
This is expected because the  KL
timescale decreases with semi-major axes
as $a^{-3/2}$ (see Equation \ref{eq:tau}).
However, we point out from panel a that for a given
semi-major axis the disruptions happen at many 
different times, not just $\tau_{\mbox{\tiny{KL}}}$. 
This is because the binary is eccentric 
($e_{\rm b}=0.5$) and we expect
the dynamics of the disk to be affected by the octupole-level
eccentricity modulations of the
KL mechanism\footnote{
Note
that the maximum eccentricity with $e_{\rm b}=0$
is $e_{\rm max}=(1-5/3\cos^2 i_{\rm b})^{1/2}=0.975$, which
it does not lead to disruptions since
$a(1-e_{\rm max})>0.3$ AU.}
(see \citealt{naoz16} for a recent 
review), which lead to extreme eccentricities
on timescales longer than $\tau_{\rm KL}$.

The strength of these octupole-level
perturbations (relative to the quadrupole-level) 
is proportional to $\epsilon_{\rm oct}=e_{\rm b}a/a_{\rm b}$, 
which implies that the planetesimals with smaller semi-major axes
are expected to be less affected by the octupole
modulations.
Consistently, we observe from panels a and b that 
the number of disruptions decreases from $\gtrsim90\%$ 
at $a\lesssim20$ AU to nearly zero for $a\lesssim14$ AU.
Overall, $70\%$ of the planetesimals are tidally disrupted
(see panel b of Figure \ref{fig:pop_dis}) . 

Similarly, the timescale is of these octupole-level oscillations is
$\sim \epsilon_{\rm oct}^{-1/2}\tau_{\mbox{\tiny{KL}}}\propto a^{-2}$
\citep{anto15}.
Therefore, the small bodies are disrupted after
several secular timescales $\tau_{\mbox{\tiny{KL}}}$
giving rise to a wide range disruptions times
for a fixed semi-major axis.
From panel c we observe that the disruption
times have a wide distribution in the range
of $\sim0.05-10$ Gyr.
The distribution peaks at $\sim0.1$ Gyr  because
the first modulation of the octupole is driving 
most of the planetesimals to extremely large 
eccentricities around this time.

Beyond this peak, the distribution flattens in log of the time,
$dN/d\log(t)\sim \mbox{cst.}$, meaning that 
it decays as $\propto 1/t$ at late times.
This slow decay is due to extra octupole-level modulations of the planetesimals
that survived the first one high eccentricity phase,
which happens preferentially for smaller values of $a$, as
expected.

In summary, most  of the planetesimal disk
($\sim70\%$ of the mass) is tidally disrupted in a wide range of 
timescales due to both the long-term octupole-level perturbations 
and the large radial extent of the disk.
The rate of disruption events decays slowly as 
$\propto1/t$ at late times.

\section{Discussion}
\label{sec:discussion}
 
 We have proposed a new mechanism to explain the 
observed pollution in WDs through the 
 tidal disruption of planetesimals orbiting these stars.
We propose that a planetary system (one or several planets 
inside $\sim2-5$ AU) shield the planetesimals
orbits against the KL  mechanism due a distant stellar
companion. 
Once the planetary system is engulfed during
the late stages
of stellar evolution (e.g., the AGB phase), the orbits of
the planetesimals become (secularly) 
unstable, leading to extreme eccentricities 
($e\gtrsim0.999$) and, therefore, to tidal disruptions.

 This mechanism has the following properties:
 \begin{enumerate}
 \item {\it pollution takes place for WDs of all cooling ages}. 
 This property, required by the observational
 evidence, is inherent to the nature of the 
eccentric KL  mechanism, which leads to the excitation into extreme 
eccentricities over very long timescales. 
\item {\it It provides a steady flux of tidally disrupted rocky bodies}. 
This property is due the large radial extent of the planetesimal disk,
widely spanning different disruption timescales. 
Each part of the disk has a disruption
timescale  $\propto a^{-2}$ (the
 eccentric KL timescale). Furthermore,
 disruptions 
 can happen after multiple eccentric KL cycles.
 This property is required in order  to have a non-negligible
  probability of observing the
 metals in the WDs atmosphere, since both the accretion and settling
 timescales are much shorter than the cooling ages of the
 observed systems. We quantify this probability in \S\ref{sec:fpoll}
 (see Equation [\ref{eq:fpoll}]).
 \item {\it The planetesimal disk can have low surface densities and be long-lived}.
 Since the planetesimal disk can have a large radial extent ($\Delta a/a\sim3.5$ in our
 example in Figure \ref{fig:pop_dis}) and most of the disk can disrupted
  ($\sim0.7$ of the mass), our mechanism can explain the
  observed accretion rates even for low surface densities
  (see constraints in \S \ref{sec:acc}). These
  low surface density disks can live for longer timescales
avoiding grinding down to dust, which would be
easily blown out during the RG and AGB phases
by radiation pressure (e.g., \citealt{BW10}).
 \end{enumerate}
 
Regarding point 1 above, we note that a similar idea, 
relying on long-timescale secular instabilities to explain the pollution of the 
oldest WDs have been presented by \citet{BV15} and
\citet{HP16}. 
These models, however, have not yet shown 
to provide the required steady flux of disrupting rocky bodies (point 2 above), 
nor consider the shielding effect against external perturbations that planetary 
systems would have during the MS phase of the host star.


In what follows, we discuss further constraints on our model
from both observations and theoretical expectations.

\subsection{Constraints on the
disk mass $M_{\rm disk}$}
\label{sec:mdisk}

Disks with high enough masses can 
quench our mechanism by 
(at least) the following two reasons:
(1)  pericenter precession rate due to self-gravity
limits the KL mechanism;
(2) dynamically hot disks become highly collisional,
likely shattering the small bodies.

The pericenter precession timescale 
of a planetesimal at 10 AU (the inner
edge of our disk)
due to the self-gravity of the disk
has been estimated by \citet{batygin11}
to be (see also \citealt{R13,TTMR15}):
\ba
\tau_{\rm s-g}\sim 13 \mbox{ Myr} \left(\frac{M_\oplus}
{M_{\rm disk}}\right).
\label{eq:tau_sg}
\ea
This timescale should be compared to the 
timescale of pericenter precession due to the external companion, $\tau_{\mbox{\tiny{KL}}}$ (Equation~\ref{eq:tau}). For the example with $a_{\rm b}=1300$ AU
and $a\sim10-35$ AU,  $\tau_{\rm KL}\sim4-24$ Myr, implying that,
for the KL mechanism to operate, we 
require $M_{\rm disk}\lesssim0.5M_\oplus$.
However, when $\tau_{\rm s-g}\gtrsim\tau_{\mbox{\tiny{KL}}}$,
the maximum eccentricity reached by the KL
mechanism is still reduced \citep[e.g.][]{liu15}
and thus the planetesimals do 
not reach star-grazing orbits.
We checked the effect of disk self-gravity 
by adding a pericenter precession
term to equations of motion with a rate
$\sim1/\tau_{\rm s-g}$, finding that a range of maximum
disk masses in $\sim0.01-0.1M_\oplus$
still allows for numerous tidal disruptions.

After KL oscillations stir up the planetesimal 
disk, the relative velocities between the planetesimals 
are high ($e\sim i\sim1$), and the collisions 
between planetesimals can be highly disruptive.
\citet{HT10} have estimated that a dynamically hot
planetesimal disk at $\sim10$ AU can survive 
the disruptive collisions or the gravitational scattering events
for Gyr timescales if
$M_{\rm disk}\lesssim M_\oplus$.
However, such disks should be comprised of 
$\lesssim10^4$ small bodies, each with at least 
a Ceres mass of $\sim10^{24}\rm{g}\sim10^{-4}M_\oplus$.
These constraints relax dramatically for disks 
at larger semi-major axes. For example, 
at 100 AU, the disk can be as massive 
100 $M_\oplus$ and allow for body numbers of
up to $10^{10}$ and individual masses
of $10^{19}$ g.
In our model, a disk with radial extent of
$\sim10-40$ AU and a mass of
$M_{\rm disk}\lesssim M_\oplus$ should survive
in the long term.

In conclusion, our mechanism is expected to  operate
in disks with masses of $\lesssim0.1M_\oplus$. 
Higher masses are expected to either quench 
the KL oscillations or disrupt the disk via collisions
between the planetesimals.

\begin{figure*}
   \centering
  \includegraphics[width=0.99\textwidth]{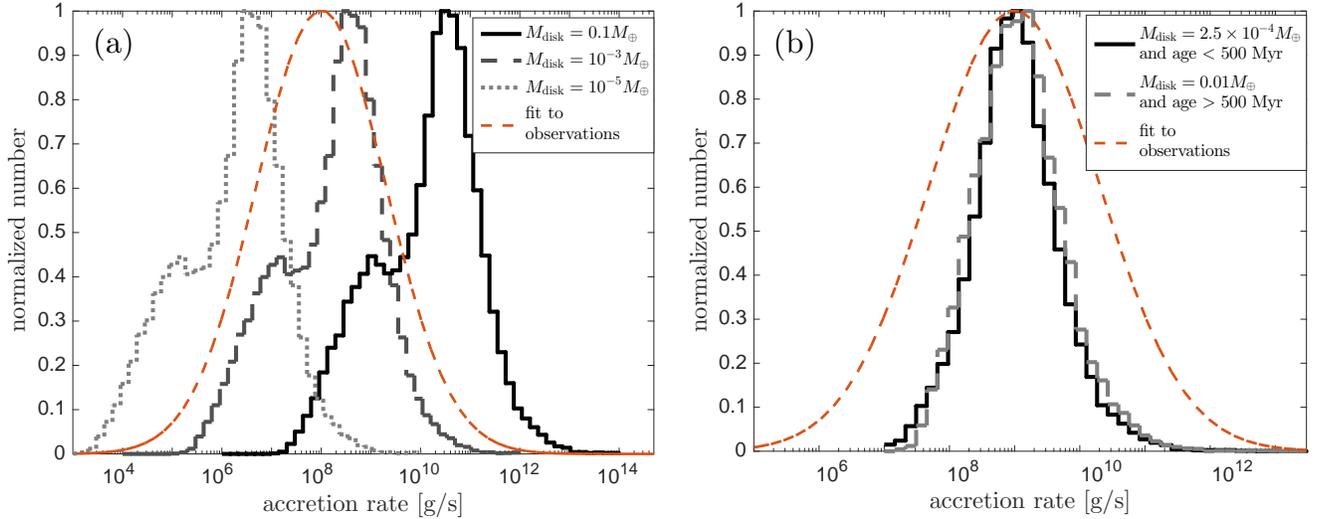}
    \caption{Accretion rates derived from the number of 
  small bodies disrupted in our
  model in Figure \ref{fig:pop_dis} for different
  disk masses $M_{\rm disk}$ and cooling ages.
  We assume that all the particles have the same mass
  and that $f_{\rm acc}=1$ in Equation (\ref{eq:Macc}), i.e., all the
  mass in tidally disrupted particles is
  accreted by the WD.
  {\it Panel a:} we set $M_{\rm disk}=0.1M_\oplus$ (solid black line), which is
  the largest mass allowed by the precession due to the self-gravity of
  the disk, and also show lower values of $M_{\rm disk}=10^{-3}M_\oplus$ 
  (dark gray dashed line) and  $M_{\rm disk}=10^{-5}M_\oplus$ (light grey
dotted line).  A fit to the observations taken from \citet{wyatt14}
is shown in red dashed line.
  {\it Panel b:}   we divide the accretion rates in panel a 
  in two different populations: cooling  ages $<500$ Myr  with
     $M_{\rm disk}=2.4\times10^{-4}M_\oplus$ 
 (solid black line) and  cooling ages $>500$ Myr 
   $M_{\rm disk}=0.01M_\oplus$ (gray dashed line).
   The disk masses in each population are set to 
   reproduce the observed peak in the observations
   (red dashed line).}
\label{fig:hist_acc}
\end{figure*}

\subsection{Accretion rates}
\label{sec:acc}

Having constrained the initial mass of the disk
to $M_{\rm disk}\lesssim0.1M_\oplus$, we can
estimate the maximum accretion rate 
of planetesimals predicted by our model as a
function of age.
For simplicity, we shall assume that the size 
distribution of the planetesimals does not change 
with semi-major axes.

The total mass accreted by the WD can estimated
as:
\ba
M_{\rm acc}=f_{\rm td}\cdot f_{\rm acc}\cdot M_{\rm disk}
\label{eq:Macc}
\ea
where $f_{\rm td}$ is the fraction of tidally disrupted objects 
and $f_{\rm acc}$ is the mass fraction of disrupted 
particles that is accreted 
and reaches the WD's surface.
We estimate the former fraction 
directly from our simulation (see panel
b of Figure \ref{fig:pop_dis}) to be $f_{\rm td}\simeq 0.7$.
The latter fraction is largely unconstrained. 

The accretion rate can be calculated as
\ba
\frac{dM_{\rm acc}}{dt}=\frac{df_{\rm td}}{dt}
\cdot f_{\rm acc}\cdot M_{\rm disk},
\ea
where $df_{\rm td}/dt$ is shown in 
panel c of Figure \ref{fig:pop_dis}
(normalized by the maximum rate).
From this figure we point out that the frequency
of disrupted bodies and, therefore the accretion
rate, is highest at $\sim50-300$ Myr and it flattens 
(in log space, i.e., decays as $\propto1/t$)
at $\gtrsim500$ Myr. 

In Figure \ref{fig:hist_acc}, we show the accretion rate
derived from our calculations by computing the
time between tidal disruptions and assuming that
all the bodies have the same mass.
We set $f_{\rm acc}=1$ and
quote values of the disk mass $M_{\rm disk}$.

In panel a, we show the accretion rate for different disk masses in a 
range of $M_{\rm disk}=10^{-5}-10^{-1}M_\oplus$, where
the upper limit is roughly the maximum mass
imposed by precession due to disk self-gravity (see \S\ref{sec:mdisk}).
The distribution of accretion for $M_{\rm disk}=0.1M_\oplus$
(solid black line) peaks 
at $\sim3\times10^{10}~\mbox{g}~\mbox{s}^{-1}$, while
there is a secondary bump at $\sim10^{9}~\mbox{g}~\mbox{s}^{-1}$.
This bi-modality is due to the accretion rate change
at early and late times in the evolution of the WD
(see panel c of Figure \ref{fig:pop_dis}).
By splitting the sample
into early stages (cooling ages $<500$ Myr) and late stages (cooling ages $>500$ Myr),
we obtain two symmetric (log-normal) distributions.
For reference, we show
a fit to the observations using a 
log-normal distribution with $\mu=8$ and $\sigma=1.3$
taken from \citet{wyatt14}.
In panel b, we show the accretion rates at early and
late stages for disk masses of 
$M_{\rm disk}=2.5\times10^{-4}M_\oplus$
and $M_{\rm disk}=0.01M_\oplus$, both coinciding with the observed
peak at $10^{8}~\mbox{g}~\mbox{s}^{-1}$.

We also note from the panel b that 
our model predicts a smaller dispersion 
of the accretion rates relative to the observations. 
In reality, we expect that for an ensemble of 
systems our predicted accretion rates 
should broaden significantly by considering 
a distribution of disk masses and 
particle masses, as well different orbital separations
of the binary companion. This calculation
is beyond the scope of this paper and it will be
worth studying in a future work.

In summary, we find that disk masses in
the range $M_{\rm disk}=10^{-4}-10^{-2}M_\oplus$
produce accretion rates consistent with observations.
These disk masses are small
enough that this mechanism is not expected 
to be limited by either by disk self-gravity 
or the collisional destruction of small
bodies (see \S\ref{sec:mdisk}).
The accretion rate is expected to peak at $\sim10^8$ yr
and decays as $\propto1/t$ at late times.

\subsection{Estimate of the pollution rate}
\label{sec:fpoll}

Here we estimate the rate at which this 
mechanism might contribute to
the observed pollution of WDs and assess
whether it can account the high observed
rates of $\sim25-50\%$.

The fraction of WDs for which our mechanism
can contribute to the observed levels of 
pollution can be estimated as:
\ba
f_{\rm poll}\equiv  f_T\cdot f_{\rm KL}\cdot f_b\cdot f_p,
\label{eq:fpoll}
\ea
where $f_T$ is the fractional time in the WD's 
cooling age during which the
rocky material is being supplied,
$f_{\rm KL}$ is the fraction
of systems that leading to planetesimal disruptions
due to the eccentric KL mechanism,
$f_b$ is the fraction of stars with wide ($a_b>100$ AU) 
binary companions, and
$f_p$ is the fraction of 
planetary systems with planets within $\sim3$ AU and
planetesimal disks during the main sequence
of A and F stars.

We can optimistically estimate that $f_T\sim1$ since
the proposed system provides a steady delivery of asteroids into
disrupting orbits (see panel c in Figure
\ref{fig:pop_dis}). In reality, we have to account for the
finite number of planetesimals in the system which would lead to
discontinuous events of tidal disruptions. However, this effect
is compensated by the finite timescales involved in the
circularization of the disrupted material and subsequent accretion
onto the WD's atmosphere (see \citealt{veras16}).
We emphasize that one of the virtues of this new mechanism is that
$f_T\sim1$, which is not the case in many others models found in
the literature.

 The fraction of systems for which the planetesimals 
 can be tidally disrupted due the eccentric KL mechanism
 $f_{\rm KL}$ can be estimated from 
previous studies in the context of planet disruptions in main
sequence stars  (e.g., \citealt{naoz12,petro15,ASL15,MLL16}).
These studies find that for a population of wide 
binaries ($a_{\rm b}\gtrsim100$ AU) with isotropic inclinations and 
a thermal eccentricity distribution\footnote{\citet{TK15} find 
that that the eccentricity distribution for solar-type wide 
binaries ($a_{\rm b}>50$ AU) 
is $f(e_{\rm b})=1.2e_{\rm b}+0.4$, while
$f(e_{\rm b})=2e_{\rm b}$ \citet{K07} for very wide binaries
($a_{\rm b}\gtrsim1000$ AU)} roughly up to $\sim30\%$  of them
can lead to planetary tidal disruptions (assuming no tidal circularization).
This estimate is consistent with the observation that $5/17\sim30\%$ of the
WDs with wide binary companions are metal-polluted
\citep{Z14}. Thus, $f_{\rm KL}\simeq0.3$.

The fraction of A and F stars (progenitors of most
WDs) with binary companions has been measured to be
 $\sim70\%$ \citep{K07,peter12,rosa12}, while the semi-major
 axis distribution follows a log-normal distribution
 peaked at $\sim300$ AU \citep{rosa12}, which is significantly
 wider than the peak of Solar-type stars at $\sim40$ AU
 \citep{ragha10}. Using the semi-major axis distribution
 from \citet{rosa12}, we find that $\sim70\%$ of the binaries
 have $a_{\rm b}>100$ AU, implying that the fraction of A stars
 with wide binary companions is $\sim0.7\times0.7\sim0.5$.
Thus, we estimate that the fraction of
A stars with wide binary companions
is $f_b\sim0.5$.

The fraction of A and F stars with planetary systems
and outer small rocky bodies $f_p$ can be estimated  --conservatively--
from the occurrence of gas giant planets (assuming they all
have small rocky bodies). 
The RV planet searches find that $\gtrsim10\%$
of Solar-type stars have gas giant planets (e.g., \citealt{WF15}),
while the occurrence seems increase linearly with the host star 
mass, reaching $\sim30\%\pm15\%$ for A-stars with 2$M_\odot$ stars
\citep{JJ10}.
Lower-mass planets around Solar-type stars are more
abundant than giants, and their
occurrence, which is not well constrained at AU separations and for A and F stars,
is roughly 0.5. 
The extrapolation of these results suggest that the occurrence of planets
in A and F stars can be as high unity. Thus, we use $f_p\sim0.3-1$.	

Putting these numbers together, we estimate that the
fraction of polluted WDs that can be explained by our mechanism is
$f_{\rm poll}=  f_T\cdot f_{\rm KL}\cdot f_b\cdot f_p\sim 1\times0.3\times0.5\times(0.3-1)\sim
0.05-0.15$. 
This result implies that our mechanism can explain 
a significant portion, but not all, of the observed polluted WDs, 
which  amount to fraction of $f_{\rm poll,obs}\gtrsim0.25$.

In summary, our mechanism can only explain a fraction, although  
still significant (up to $50\%$), of the observed polluted WDs. 
Complete searches for companions, for which 
GAIA will play a crucial role,  might shed light on the significance of 
this new model to explain the pollution of WDs.

\subsection{Outer companions in polluted white dwarfs}

The engulfment-aided KL mechanism to explain WD pollution requires the presence of a massive body
in a wide orbit
(either a stellar companion or a planet). 
We note that both the distance and the mass of the distant
perturber enter into the calculation mainly through  
the timescale of the gravitational perturbations in the form of
the dimensionless parameter $\epsilon\out\propto M\out/a\out^3$
(see Equations [\ref{eq:varpi_out}] and [\ref{eq:eps_out}]). 
Thus, the evolution of the system with a Solar-mass companion 
will be very similar to that with a Jupiter-mass companion but 10 times closer,
providing an alternative version of the mechanism that may
increment the explained pollution rate by accommodating different outer
companions.

We note that the presence of a low-mass main sequence star like in the 
example of Figure \ref{fig:tau_a} would most likely be detected
had the WD been observed in the first place. However, the fraction of 
polluted WDs with wide main sequence star companions is small
(e.g., \citealt{Z14}). 
This implies that either the model presented here
can only explain a subset of the polluted WDs, or that the current sample 
of outer companions of polluted WDs is largely incomplete.
We briefly discuss the latter alternative, namely,  
other types of companions might still escape detection and might contribute 
to the incompleteness of the current sample of companions.

First, outer planetary companions and brown dwarfs at $\sim10-100$ AU 
distances can remain undetected (e.g., \citealt{farihi05,debes11,day13}). 
Also, these outer companions can drive eccentricities to nearly 
unity values not only by the KL mechanism
(for which large inclinations respect to the 
planetesimal disk are required), but also by either 
a nearly coplanar and eccentric body \citep{li14,petro15b}
or secular chaotic diffusion due to two or more eccentric and/or 
inclined bodies \citep{LW11,WL11,batygin15}.

Second, stellar-mass companions such as 
other fainter WDs or neutron stars
and black holes can also escape detection.
One intriguing observational puzzle
is the mismatch between
the measured binary fraction 
of WDs \citep[$\sim30\%$;][]{farihi05,holberg16} and
that of their progenitors 
($\sim 70-100\%$, e.g., \citealt{K07}).
As noted by \citet{ferr12}, a relatively flat mass ratio distribution 
for the progenitors of WDs gives a better fit the mass distribution of the detected WDs companions, typically M-dwarfs.
However, this same distribution also implies that 
$\sim30\%$ of the WDs should be in double WD systems,
most of them ``hiding" as singles.
Similarly, based on the completeness of the sample of 
A-stars in \citet{rosa12}, \citet{KK16} argued that 
$\sim10\%$ of these WD progenitors are likely to host 
undetected companions that will become WDs within the age 
of our galaxy. In this picture, current catalogs are 
typically missing the fainter WDs in the WD-WD system.

\subsection{Effects ignored and 
simplifications}
\label{sec:extra}

We discuss some of the relevant effects ignored in this
work that might change the dynamics of the system.

 \paragraph{Extra planets in the system}
For simplicity,  we have
considered a planetary systems with only one planet. 
We expect our results not to be significantly altered
if extra planets within $\sim1-5$ AU are present in the system (those would also be engulfed).
In particular, the extra planets enhance the precession
rate of the small bodies in the disk and shield the bodies from 
outer perturbations at even larger distances (i.e., the Laplace radius in Equation [\ref{eq:rl}]
increases\footnote{By adding $i=1,..,N$ bodies
with masses $m_i$ and semi-major axes $a_i$
we just replace $m_{\rm p}a_{\rm p}^2$ 
by $m_{\rm p}a_{\rm p}^2+\sum_{i=1}^Nm_ia_i^2$
in the Laplace radius
in Equation (\ref{eq:rl}).}).

If one or more distant enough planets 
do survive the AGB phase, these
can quench the secular instabilities and our mechanism
would not operate (at least in its cleanest version described
here). However, the surviving planets themselves could be susceptible
to secular perturbations due to the distant perturber.
This effect can, in principle, destabilize a 
planetary system and/or excite the eccentricities of the planets,
and such processes in turn can lead to the tidal disruption of small
bodies by the WD.

 \paragraph{Mass loss and galactic tides on the
very wide binaries}  
For wide enough binaries ($\sim10^4$ AU), 
mass loss can happen on timescales that are not too short compared
to the orbital timescales and the response of the orbital
elements would be different \citep{veras11,BV15} from what is 
described in Section~\ref{sec:trigger}. In general, this implies that either 
the companion becomes more eccentric (or even unbound), which
can enhance the eccentric KL mechanism (e.g., \citealt{naoz16}).
In addition, these wide binaries will be affected by galactic tides
and have their angular momentum altered, which again has the potential
of enhancing the secular interactions.

\paragraph{Stellar evolution}
We have modeled the planetary engulfment by shrinking
the orbit in a prescribed way (see Equation \ref{eq:aplanet})
to see how the orbital elements of the planetesimal
respond to a gradual planet  removal. 
In reality, the process is much more complicated and 
the radius of the star during the AGB can undergo
pulsations, while tides in the star lead to planet
inspiral.
These effects have been modeled in detail by 
\citet{MV12} and \citet{villa14}.
Their results indicate that the engulfment
generally happens on timescales of $\sim0.1$ Myr. 
This timescale is shorter than the shortest secular 
timescale\footnote{The shortest 
secular timescale can be obtained from 
evaluating $\tau_{\mbox{\tiny{KL}}}$ in Equation (\ref{eq:tau}) 
at a semi-major axis and period of a planetesimal 
located at the Laplace
radius in Equation (\ref{eq:rl}).} $\tau_{\mbox{\tiny{KL}}}$ of planetesimals
during engulfment for binaries with
$a_{\rm b}\gtrsim100$ AU. Therefore, we expect
our model to operate for these wide binaries regardless
of the details of planet engulfment.

In a future work, we will calculate the effects of mass 
loss and the planetary engulfment using realistic stellar 
evolution models and incorporate these in 
a population synthesis model.
By doing this, we will be able to better address
the significance of our model.

\section{Conclusions}

We have studied a new mechanism to explain the 
observed metal pollution in white dwarfs through the tidal 
disruption of small rocky bodies in a planetesimal
disk.
We propose that one or several planets can shield a
planetesimal disk against the KL mechanism due a 
distant binary companion.
Once the host star evolves off the main sequence to become
a WD, these planets can be engulfed
(most likely during the AGB phase), thus triggering the
KL mechanism, and leading to the tidal
disruption of the rocky bodies in the planetesimal 
disk.

We have shown that this mechanism can account for the 
observed accretion rates for WDs with all cooling
ages provided that the disks have masses $\sim10^{-4}-10^{-2}M_\oplus$.
Our model allows for planetesimal disks with large radial extents, and as a consequence,
it  presents the following advantages compared 
to other models: 
\begin{itemize}
\item it provides a steady supply of material (each part of the
disk has a different and long disruption timescale), enhancing the
probability of observing the pollution of WD atmospheres;
\item it allows for low-density surface disks, which can 
survive internal disruptive collisions over long timescales.

\end{itemize}

This mechanism is only triggered after the host star has left the main sequence, providing a self-consistent explanation as to why the KL mechanism does not act on the planetesimal disk for the prior few Gyrs.
Our estimates indicate that this model can account 
for a significant fraction of the polluted WDs. Complete searches 
for companions of WDs might shed light on the significance
of our proposal.

\acknowledgements{
We are grateful to Daniel Tamayo, 
Dimitri Veras, Dong Lai, Norm Murray, Nicholas Stone, Roman Rafikov, 
and Yanqin Wu for enlightening discussions.
C.P. acknowledges 
support from the Gruber Foundation Fellowship.}

\appendix
\section{Secular equations of motion}

Consider a mass-less planetesimal orbiting the center of mass
of a star with mass $M_s$ and an inner planet
with mass $m_p$ (total and reduced masses $M\in=M_{\rm s}+m_{\rm p}$ and
$\mu\in=M_{\rm s}M_{\rm p}/M\in$, respectively)
with semi-major axis $a$ with eccentricity $e$. The instantaneous
Keplerian orbit of this body oriented in space by the eccentricity
vector $\evec$ and the dimensionless specific angular momentum vector
$\jvec=\sqrt{1-e^2}~\hat{\bf{j}}$.
The host star is also a member of a wide binary $M_b$.
Thus, the asteroid orbit is perturbed by two non-Keplerian potentials, which
we call $\langle{\Phi\in\rangle}$ (due to the star-planet pair, of orbital parameters
$a\in$ and $e\in$) and $\langle{\Phi\out\rangle}$ (due to the distant
stellar binary, of orbital parameters $a\out=a_b$ and $e\out=e_b$).
The  non-Keplerian potential averaged over all the orbits
to octupole order in  the semi-major axis ratio
and including a term
to describe the precession due to General Relativity (GR) is:
\begin{equation}
\label{eq:potential_av}
\Phi=\langle{{\Phi_{\mathrm{in,quad}}}}\rangle
+\langle{{\Phi\in},\oct}\rangle\\
+\langle{{\Phi_\mathrm{out,quad}}}\rangle
+\langle{{\Phi\out},\oct}\rangle
+\langle{\Phi_\mathrm{cross,Oct}}\rangle
+\langle{\Phi_\mathrm{GR}}\rangle,
\end{equation}
where $\langle{\Phi_\mathrm{cross,Oct}}\rangle$ is a cross-term
coupling the planetesimal's orbit to both the inner and 
outer orbits \citep{ham15}.
Following the notation of \citet{mun15b}, we write these potentials as:
\ba
\label{eq:potential_inner_av}
\langle{{\Phi_\mathrm{in,quad}}}\rangle(\evec,\jvec)&=&-\frac{1}{4} \frac{\mathcal{G}M\in}{a}
\epsilon\in{(1-e^2)^{-5/2}}
\Big[(1-6e\in^2)(1-e^2)-3(1-e\in^2)(\nvec\in\cdot\jvec)^2
+15e\in^2(\uvec\in\cdot\jvec)^2\Big],\\
\label{eq:potential_outer_av}
\langle{{\Phi_\mathrm{out,quad}}}\rangle(\evec,\jvec)&=&-\frac{1}{8}\frac{\mathcal{G}M\in}{a}
\epsilon\out (1-e\out^2)^{-3/2}
\Big[1-6e^2-3(\nvec\out\cdot\jvec)^2+
15(\nvec\out\cdot\evec)^2\Big],\\
\langle{{\Phi\in},\oct}\rangle(\evec,\jvec)&=&-\frac{15}{32} \frac{\mathcal{G}M\in}{a}
\epsilon{\in}\,\epsilon{\in}{,\oct}{(1-e^2)^{-7/2}}
\times\Bigg\{(\evec\cdot\uvec\in)\Big[(8e\in^2-1)(1-e^2)-35e\in^2(\jvec\cdot\uvec\in)^2
\\
&&~~~~~~~~~~~~~~~~~~~~~~~~~~~~~+5(1-e\in^2)(\jvec\cdot\nvec\in)^2\Big]
+10(1-e\in^2)
(\evec\cdot\nvec\in)
(\jvec\cdot\uvec\in)
(\jvec\cdot\nvec\in)\Bigg\},\\
\label{eq:potential_outer_av_oct}
\langle{{\Phi\out},\oct}\rangle(\evec,\jvec)&=&
-\frac{15}{64}\frac{\mathcal{G}M\in}{a}\epsilon\out\,\epsilon{\out}{,\oct}
\Bigg\{(\evec\cdot\uvec\out)\Big[8e^2-1-35(\evec\cdot\nvec\out)^2
+5(\jvec\cdot\nvec\out)^2\Big]
+10(\evec\cdot\nvec\out)
(\jvec\cdot\uvec\out)(\jvec\cdot\nvec\out)\Bigg\},\nonumber \\
\\
\langle{\Phi_\mathrm{cross,Oct}}\rangle(\evec,\jvec)&=&-\frac{9}{32}\frac{\mathcal{G}M\in}{a}
\epsilon\in\epsilon\out\epsilon{\out}{,\oct}
\Bigg\{2(1-e\in^2)(\evec\cdot\nvec\in)(\uvec\out\cdot\nvec\in)\Big[4-5(\uvec\in\cdot\nvec\out)^2\Big]
\nonumber\\
& &-10(\uvec\in\cdot\uvec\out)(\uvec\in\cdot\nvec\out)\Big[(1+e\in^2)(\evec\cdot\nvec\out)
-(1-e\in^2)(\evec\cdot\nvec\in)(\nvec\in\cdot\nvec\out)\Big]
\nonumber\\
&&+(\evec\cdot\uvec\out)\Big[-(1-6e\in^2)-10(1-e\in^2)(\uvec\out\cdot\nvec\in)^2
-5(\uvec\in\cdot\nvec\out)^2\Big(5e\in^2-2(1-e\in^2)(\uvec\out\cdot\nvec\in)^2\Big)
\nonumber\\
&&-20(1-e\in^2)(\uvec\in\cdot\uvec\out)(\uvec\in\cdot\nvec\out)(\uvec\out\cdot\nvec\in)(\nvec\in\cdot\nvec\out)
+5(1-e\in^2)(\nvec\in\cdot\nvec\out)^2
\nonumber\\
&&-10(1-e\in^2)(\uvec\in\cdot\uvec\out)^2\Big(1-(\nvec\in\cdot\nvec\out)^2\Big)\Big]
+10(\nvec\in\cdot\evec)\Big[(1-e\in^2)(\uvec\in\cdot\uvec\out)(\uvec\out\cdot\nvec\in)(\nvec\in\cdot\nvec\out)
\nonumber\\
&&+(\uvec\in\cdot\uvec\out)\Big(1-(1-e\in^2)(\nvec\in\cdot\nvec\out)^2\Big)\Big]
\Bigg\}~~,\\
\label{eq:potential_av_gr}
\langle{\Phi_\mathrm{GR}}\rangle(\evec)&=&3\left(\frac{\mathcal{G}M\in}{a}\right)^2\frac{1}{c^2\sqrt{1-e^2}}~~.
\ea
where the unit vectors $(\uvec\in,\nvec\in)$ and $(\uvec\out,\nvec\out)$
define the orientation of the eccentricity and the angular momentum vectors for
the inner and outer orbits, respectively, and we have defined
\ba
\epsilon\in&\equiv& \frac{1}{2}\left(\frac{\mu\in}{M\in}\right)\left(\frac{a\in}{a}\right)^2~~,\\
\epsilon\out&\equiv&  \frac{1}{(1-e\out^2)^{3/2}}\left(\frac{M\out}{M\in}\right)\left(\frac{a}{a\out}\right)^3~~,
\label{eq:eps_out}\\
\epsilon{\out}{,\oct}&\equiv&\left(\frac{a}{a\out}\right)\frac{e\out}{1-e\out^2},\\
\epsilon{\in}{,\oct}&\equiv& e\in\left(\frac{a\in}{a}\right)\sqrt{1-4\mu\in/M\in}~~,
\ea
which characterize the relative strength of the perturbing potentials
 for a given $a$.

In our calculations, the planet remains in a low-eccentricity orbit ($e\in\ll1$), which 
implies that $\epsilon{\in}{,\oct}\ll1$. Therefore, the amplitude
of $\langle{\Phi_\mathrm{cross,Oct}}\rangle$ is much smaller than all the terms 
of the potential in Equations (\ref{eq:potential_inner_av})-(\ref{eq:potential_outer_av_oct})
and can be thus neglected.

The equations of motion for $\evec$ and $\jvec$ can be written as
\ba
\label{eq:motion_tp_j}
\frac{d\jvec}{dt}&=&\frac{1}{\sqrt{\mathcal{G}M\in a}}\Big(
\jvec\times\nabla_\jvec \Phi
+\evec\times\nabla_\evec \Phi
\Big)\\
\label{eq:motion_tp_e}
\frac{d\evec}{dt}&=&\frac{1}{\sqrt{\mathcal{G}M\in a}}\Big(
\jvec\times\nabla_\evec \Phi
+\evec\times\nabla_\jvec \Phi
\Big)
\ea

In turn, the background system of star, planet, and companion 
(which is unaffected by the presence of the massless particle)
evolves according to the secular dynamics of triple systems.
In what follows, we only write the equations of motion for the inner
star-planet pair as the outer stellar binary remains roughly fixed during the
evolution of the system (i.e., $\evec\out$ and $\jvec\out$ are roughly constant).
Similar to the case of the planetesimal's orbit, we write the 
non-Keplerian
potential of the inner system as:
\ba
\Phi\in&=&\langle{{\Phi_{\rm in-out,Quad}}}\rangle
+\langle{{\Phi{\inout}},\oct}\rangle
+\langle{\Phi{\in}_\mathrm{,GR}}\rangle,
\ea
where
\ba
\langle{{\Phi_{\rm in-out,Quad}}}\rangle(\evec\in,\jvec\in)&=&-\frac{1}{8} \frac{\mathcal{G}M\in}{a\in}
{\mu\in}\frac{\left({M\out}/{M\in}\right)\left({a\in}/{a\out}\right)^3}{(1-e\out^2)^{3/2}}\Big[(1-6e\in^2)-3(\jvec\in\cdot\nvec\out)^2+15(\evec\in\cdot\nvec\out)^2\Big]~~,\\
\langle{{\Phi\inout},\oct}\rangle(\evec\in,\jvec\in)&=&
-\frac{15}{64} \frac{\mathcal{G}M\in}{a\in}\mu\in
\left({M\out}/{M\in}\right)\left({a\in}/{a\out}\right)^4\,\sqrt{1-4\mu\in/M\in}
\frac{e\out}{(1-e\out^2)^{5/2}}
\nonumber\\
\times\Bigg\{(\evec\in\cdot\uvec\out)
&\times&
\Big[(8e\in^2-1)-35(\evec\in\cdot\nvec\out)^2+5(\jvec\in\cdot\nvec\out)^2\Big]
+10(\jvec\in\cdot\uvec\out)(\evec\in\cdot\nvec\out)(\jvec\in\cdot\nvec\out)\Bigg\}~~,\\
\langle{\Phi{\in}_\mathrm{,GR}}\rangle&=&3\mu\in\left(\frac{\mathcal{G}M\in}{a\in}\right)^2\frac{1}{c^2\sqrt{1-e\in^2}}
\ea

The equations of motion for $\evec\in$ and $\jvec\in$ can be written as
\ba
\label{eq:motion_pl_j}
\frac{d\jvec\in}{dt}&=&\frac{1}{\sqrt{\mathcal{G}M\in a\in}}\Big(
\jvec\in\times\nabla_{\jvec_{\rm in}} \Phi\in
+\evec\in\times\nabla_{\evec_{\rm in}}\Phi\in
\Big),\\
\label{eq:motion_pl_e}
\frac{d\evec\in}{dt}&=&\frac{1}{\sqrt{\mathcal{G}M\in a\in}}\Big(
\jvec\in\times\nabla_{\evec_{\rm in}} \Phi\in
+\evec\in\times\nabla_{\jvec_{\rm in}} \Phi\in
\Big).
\ea

We incorporate the effect from mass loss and engulfment
of the planet due to tides on the star as:
\ba
\frac{\dot{a}\in}{a\in}=-\frac{\dot{M}_s}{M\in}-\tau_a^{-1}\equiv \tau_{\rm ml}^{-1}
\frac{M_s}{M\in}-\tau_a^{-1},
\label{eq:ml_in}
\ea 
while we only prescribe the orbit expansion due to mass loss
for the planetesimal and outer (binary) orbits as
\ba
\frac{\dot{a}}{a}&=&\tau_{\rm ml}^{-1} ~\mbox{ and}
\label{eq:ml}\\
\frac{\dot{a}_{\rm out}}{a_{\rm out}}&=&\tau_{\rm ml}^{-1} \frac{M_s}{M_s+M_{\rm out}}
\label{eq:ml_out}
\ea
respectively.

\end{document}